\documentclass[preprint2,trackchanges]{aastex631}

\usepackage{graphicx}
\usepackage[suffix=]{epstopdf}
\usepackage{amsmath}
\usepackage{url}
\usepackage{xspace}
\usepackage{mathrsfs}
\usepackage{enumitem}
\usepackage[caption=false]{subfig}
\usepackage{float}

\begin{document}

\newcommand{\UCB}{Radio Astronomy Lab, University of California, Berkeley, 501 Campbell Hall 3411, Berkeley, CA, 94720, USA}
\newcommand{\seti}{SETI Institute, Mountain View, California}
\newcommand{\yale}{Department of Astronomy, Yale University, Steinbach Hall, 52 Hillhouse Ave, New Haven, CT 06511}
\newcommand{\uw}{Department of Astronomy, University of Washington, Box 351580, Seattle, WA 98195-1700}
\newcommand{\jbca}{Jodrell Bank Centre for Astrophysics (JBCA), Department of Physics \& Astronomy, Alan Turing Building, The University of Manchester, M13 9PL, UK}
\newcommand{\malta}{University of Malta, Institute of Space Sciences and Astronomy}

\author[0000-0002-5956-5167]{Andy Nilipour}
\affiliation{\yale}
\affiliation{\UCB}

\author[0000-0002-0637-835X]{James R. A. Davenport}
\affiliation{Department of Astronomy, University of Washington, Box 351580, Seattle, WA 98195, USA}

\author[0000-0003-4823-129X]{Steve Croft}
\affiliation{\UCB}
\affiliation{\seti}

\author[0000-0003-2828-7720]{Andrew P. V. Siemion}
\affiliation{\UCB}
\affiliation{\seti}
\affiliation{\jbca}
\affiliation{\malta}

\title{Signal Synchronization Strategies and Time Domain SETI with Gaia DR3}

\shorttitle{Time Domain SETI with Gaia DR3}
\shortauthors{Nilipour et al.}


\begin{abstract}
Spatiotemporal techniques for signal coordination with actively transmitting extraterrestrial civilizations, without the need for prior communication, can constrain technosignature searches to a significantly smaller coordinate space. With the variable star catalog from Gaia Data Release 3, we explore two related signaling strategies: the SETI Ellipsoid, and that proposed by Seto, which are both based on the synchronization of transmissions with a conspicuous astrophysical event. This dataset contains more than 10 million variable star candidates with light curves from the first three years of Gaia’s operational phase, between 2014 and 2017. Using four different historical supernovae as source events, we find that less than 0.01\% of stars in the sample have crossing times, the times at which we would expect to receive synchronized signals on Earth, within the date range of available Gaia observations. For these stars, we present a framework for technosignature analysis that searches for modulations in the variability parameters by splitting the stellar light curve at the crossing time.
 
\end{abstract}


\section{Introduction}

Searches for technosignatures, the detection of which would be indicative of extraterrestrial intelligence, must grapple with a vast, multidimensional parameter space that leaves unknown the nature of the signal, and the spatial and temporal locations of the transmission \citep{wright2018c}. Because we are unable to constantly monitor across all possible remote sensing modalities, we must select when, where, and how to conduct our observations. An advanced extraterrestrial civilization could infer this searching difficulty, and may synchronize its signal with some conspicuous astrophysical event. These source events, which should be easily visible and uncommon, act as ``Schelling'' or focal points \citep{schelling1960}, allowing for coordination despite a lack of communication between the two parties.

The distribution of stars from which a transmission synchronized with an event that would be observable on Earth forms a so-called search for extraterrestrial intelligence (SETI) Ellipsoid, and it can be used to greatly constrain the number of technosignature observation candidates. Though this framework has been described in the past \citep[e.g.,][]{makovetskii1977,lemarchand1994}, a precise application of the SETI Ellipsoid requires precise stellar positions and distances, which has only recently been made possible by Gaia. 

Gaia Early Data Release 3 \citep[Gaia EDR3;][]{gaia_edr3} contains the full astrometric solution of nearly 1.5 billion stars, which has allowed for the calculation of precise photogeometric distances \citep{bailer-jones2021}. Decreased uncertainties in stellar distances reduce the timing uncertainty of observations using the SETI Ellipsoid scheme. However, the Gaia catalog contains stars up to tens of kpc away, and at these large distances, the errors in distance translate to impractically high timing uncertainties.

\begin{figure}[t]
\centering
\includegraphics[width=3.3in]{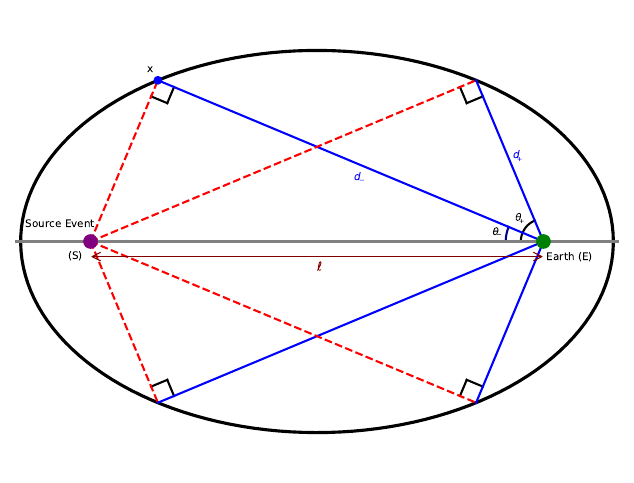}
\caption{The angles of observation for the Seto scheme. Synchronized signals from stars along the four blue lines, which correspond to two conical shells when rotated about the horizontal axis, will be observable on Earth now; these systems are the current technosignature candidates for this signaling framework. The path from the source event through the potentially transmitting civilization to Earth, through a pair of red and blue lines, forms a right angle at a point on the SETI Ellipsoid, where the technosignature candidate is located.}
\label{fig:seto_two_angle}
\end{figure}

To mitigate the effect of stellar distance uncertainties, \citet{seto2021} considers a framework in which an extraterrestrial civilization follows a certain geometric signaling scheme with a time-dependent directionality, rather than sending out an isotropic transmission at one point in time, and observers on Earth follow a closely linked receiving scheme. In this approach, shown in Figure \ref{fig:seto_two_angle}, rather than observing stars lying on an ellipsoid in Galactic Cartesian coordinates at a given point in time, we would instead observe along two concentric rings on the celestial sphere, up to a certain depth. Using this, we do not need to precisely know the distance to the candidate star, only that it is less than some upper bound, giving an advantage to the Seto scheme.

With the SETI Ellipsoid and the Seto scheme, we can assign each star an `Ellipsoid' and a `Seto' crossing time that indicates when that star should be observed according to the respective coordination framework. Scheduling future observations is possible with this methodology, as is looking through archival and publicly available data. Gaia Data Release 3 \citep[Gaia DR3;][]{gaia_dr3} contains variability analysis and type classification, along with epoch photometry, for approximately 10.5 million stars. Though these photometry data are limited, one way we can search for extraterrestrial intelligence signals is to perform independent variability analysis on the stellar light curves before and after the stars' crossing time. Such an analysis would be primarily sensitive to modulations in a periodic star’s frequency, phase, and amplitude, as well as any other statistically significant differences in light-curve properties before and after a star's crossing time, which may be a form of information transmitted by a sufficiently advanced civilization.

In this paper, we use Gaia to explore the SETI Ellipsoid and the Seto scheme as techniques for analyzing past observations in addition to prioritizing future technosignature searches.\footnote[1]{Source code available at \citet{nilipourGit}} In Section \ref{sec:gdr3} we present the sample of variable stars identified in Gaia DR3. In Section \ref{sec:elip} we discuss the geometry of, and our methods of target selection relative to, the SETI Ellipsoid, and in Section \ref{sec:seto} we do the same for the Seto scheme. In Section \ref{sec:cand} we explore our candidate sample derived from both methodologies. In Section \ref{sec:analysis} we present a novel time domain SETI approach that we apply to a subset of the periodic variables in the sample. Finally, in Section \ref{sec:conclusion} we conclude with a summary of our work, its limitations, and possibilities for future work.

\section{Gaia Data Release 3}
\label{sec:gdr3}

\begin{figure}[t]
\centering
\includegraphics[width=3in]{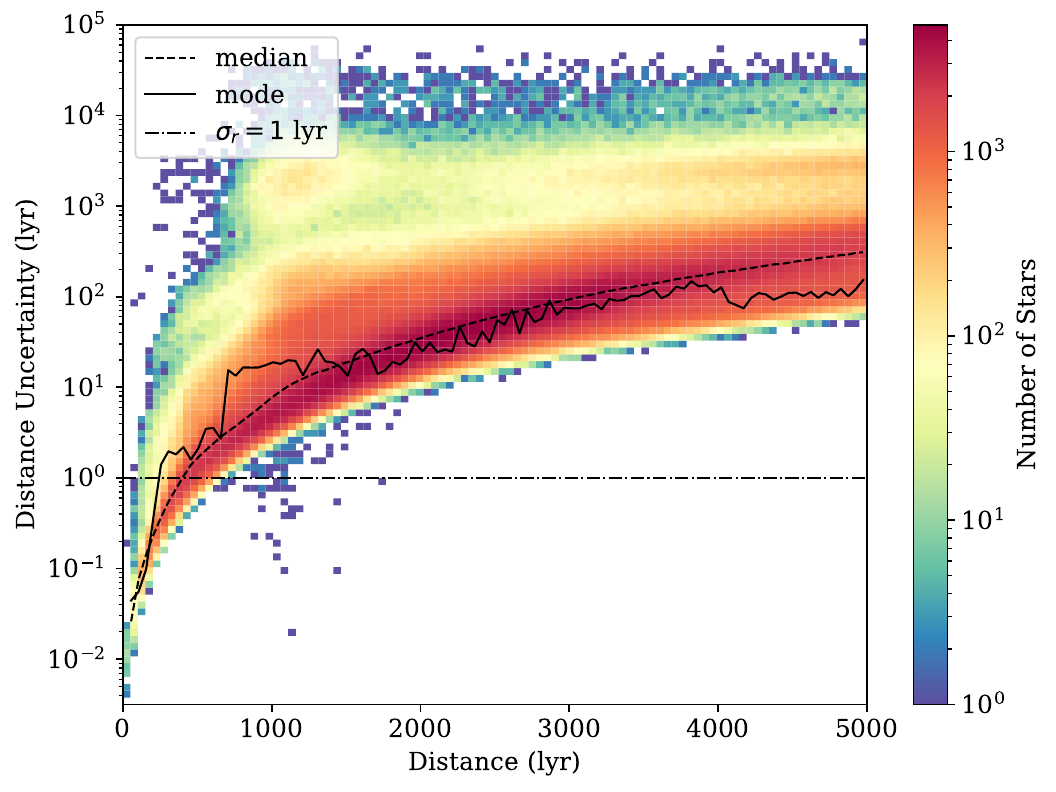}
\caption{
Two-dimensional histogram of the distance uncertainties for the variable stars in Gaia DR3, with distances less than 5000 lyr. The median and mode distance uncertainties are shown in the dashed and solid lines, respectively. Stars with a distance uncertainty less than 1 lyr are the most ideal; there are 43,237 such stars total, mostly at low distances.}
\label{fig:distance_uncertainty}
\end{figure}

Gaia DR3 \citep{gaia_dr3}, the most recent data release of the Gaia mission, complements the astrometric and photometric data of the previous EDR3 \citep{gaia_edr3} with spectra, radial velocities, astrophysical parameters, and rotational velocities for subsets of the full catalog, as well as epoch photometry and variability analysis for 10.5 million stars, which is of particular interest for this work. Each of these stars is placed into one of 24 variable classifications. Although stars that have not been classified as a known variable type by Gaia could still harbor a signal in their light curve, such as a single flux measurement with a significant deviation observed at the crossing time, the epoch photometry for these stars is not yet available, and so not considered for this analysis. However, our approach described in Section \ref{sec:analysis} can be easily applied to such stars when their light-curve data are released.

\begin{figure*}[t]
\centering
\includegraphics[width=7in]{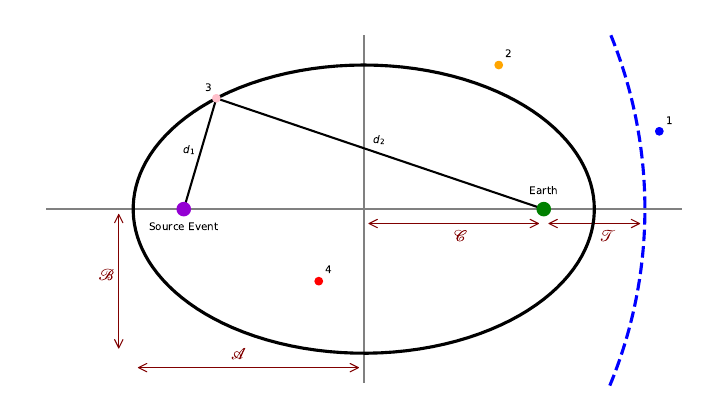}
\caption{Diagram of the SETI Ellipsoid, which has its foci at the source event (purple dot) and Earth (green dot). The blue dashed line represents the information front of the source event; stars outside this sphere (blue dot 1) have not yet observed the event. Stars outside the ellipsoid (orange dot 2) have observed the event, and if they transmitted a signal upon observation, such a signal would arrive at Earth in the future. For stars on the ellipsoid (pink dot 3), their synchronized transmission would be observable on Earth now; at any time, these are the current technosignature candidates. Signals from stars inside the ellipsoid (red dot 4) would have been received on Earth in the past.}
\label{fig:elip}
\end{figure*}

With a prior that factors in Gaia's magnitude limit and interstellar extinction, \citet{bailer-jones2021} produced photogeometric distance estimates (using the Gaia parallax, color, and apparent magnitude) for 1.35 billion stars. \citeauthor{bailer-jones2021} give the median, $r_\text{med}$, and the 16th and 84th percentiles, $r_\text{lo}$ and $r_\text{hi}$, respectively, of the photogeometric posterior. In this work, the distances used are $r_\text{med}$ and the distance uncertainties used are the symmetrized uncertainty $\sigma_r = (r_\text{hi} - r_\text{lo})/2$.

Of the 10.5 million total variable stars, 9.3 million have Bailer-Jones distances; henceforth, we refer to this set as the Gaia catalog of variable stars. As seen in  Figure \ref{fig:distance_uncertainty}, the distance uncertainties in these measurements are generally large, which makes timing uncertainty for signal synchronization via the SETI Ellipsoid, which is directly related to the stellar distance uncertainty, unfeasible. Around 0.5\% of Gaia variable stars have a distance uncertainty less than 1\,ly, with the distances to all of these stars less than 2000\,ly. 

Gaia epoch photometry has excellent long-term stability, which is useful when considering the potentially large timing uncertainties on the stellar crossing times. However, the light curves for individual stars are sparse, which makes searching for transient signals from extraterrestrial civilizations challenging. Instead, our search for signals, described in Section \ref{sec:analysis}, relies on longer-term signals involving the variability of the host star. Additionally, the epoch photometry in Gaia DR3 only includes data from a three-year period between 2014 and 2017, and thus contains less than half of the data collected to date, which limits our target selection accordingly.

\section{Target Selection via the SETI Ellipsoid}
\label{sec:elip}

\begin{figure*}[t]
\centering
\includegraphics[width=5in]{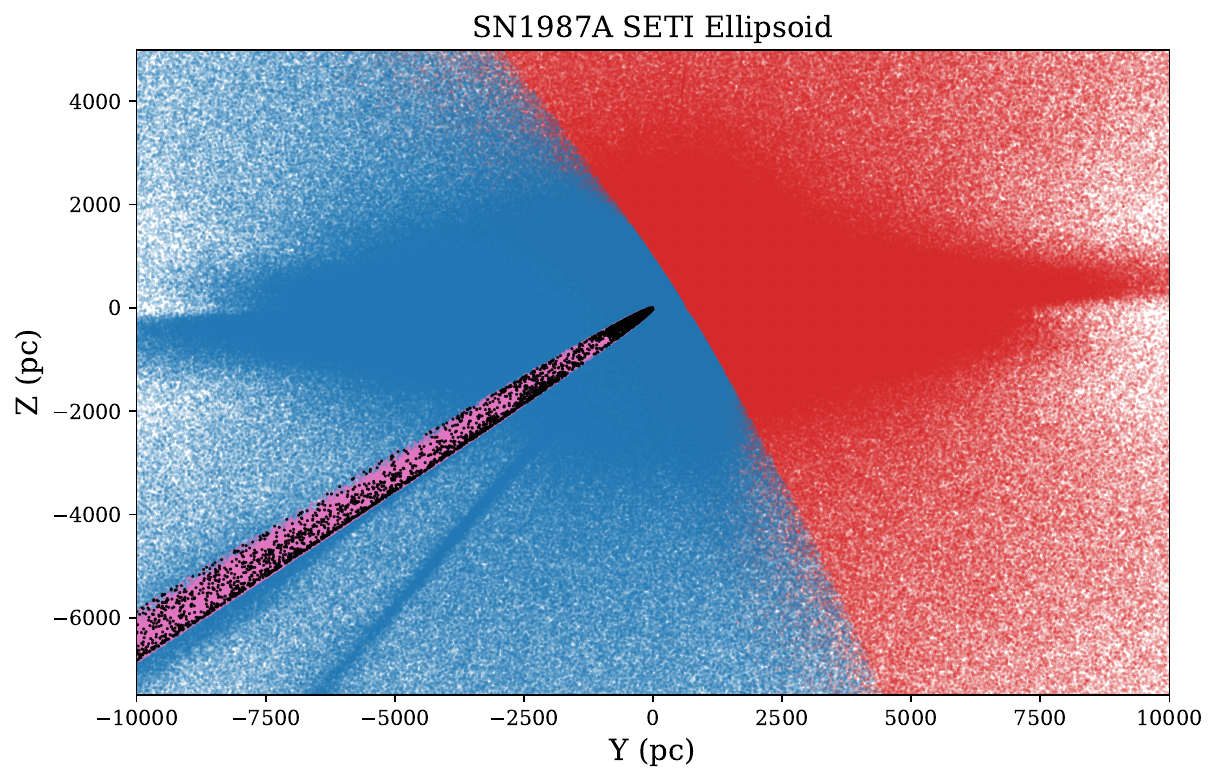}
\caption{Current SETI Ellipsoid with SN\,1987A as the source event, in galactocentric $Y$ and $Z$ coordinates. The red dots are stars that have not yet seen SN\,1987A; the blue dots have seen it, but are outside the SETI Ellipsoid. The pink dots are stars inside the SETI Ellipsoid, and the black dots are those within 0.1\,ly of the SETI Ellipsoid.}
\label{fig:1987elip}
\end{figure*}

The SETI Ellipsoid framework is based on the simplest synchronized communication system, in which an extraterrestrial civilization transmits an isotropic signal when it observes the source event, and has been previously proposed as a method to constrain the technosignature search space \citep{lemarchand1994}. The geometry of the SETI Ellipsoid is shown in Figure \ref{fig:elip} and described here.

All stars that have observed a source event are contained within the sphere centered at the event with radius
\[\mathcal{R} = \mathcal{D} + \mathscr{T}\]
where $\mathcal{D}$ is the distance from the event to Earth and $\mathscr{T} = cT$ is the current time elapsed since Earth observation of the source event ($T$) times the speed of light (c). Suppose each of these stars hosts an extraterrestrial civilization that has transmitted a signal upon its observation of the source event. The stars from which these emitted signals reach Earth at any given time lie on the SETI Ellipsoid. Formally, the ellipsoid, which has foci at Earth and the source event, is defined by
\[d_1 + d_2 = \mathcal{D} + \mathscr{T}\]
where $d_1$ is the distance from the event to the star and $d_2$ is the distance from the star to Earth. As can be seen in Figure \ref{fig:elip}, the linear eccentricity of the ellipse is
\[\mathscr{C} = \frac{\mathcal{D}}{2}\]
and the semi-major axis is
\[\mathscr{A} = \frac{\mathcal{D} + \mathscr{T}}{2}\]
So, the semi-major axis grows at half the speed of light.

Using this schematic, we can categorize all stars into four interest levels, which correspond to the labels 1--4 in Figure \ref{fig:elip}:
\begin{enumerate}[leftmargin=9mm]
\item Stars outside the sphere of radius $\mathcal{R}$, which have not yet seen the source event. Synchronized transmissions from these stars will not be observable until a minimum of $T$ years from now, which, for most events, is unreasonably far in the future.
\item Stars within the radius $\mathcal{R}$ sphere but outside the SETI Ellipsoid, from which a synchronized signal would have already been sent but not received on Earth. Signals from these stars will not be observable for a maximum of $T$ years; the stars in this category that are closer to Earth may be potential candidates for scheduling future technosignature searches.
\item Stars on the SETI Ellipsoid. Synchronized signals from these stars would be arriving at Earth now, making these ideal candidates for immediate observations.
\item Stars inside the SETI Ellipsoid, from which synchronized signals would have already been received in the past. Archival data or previous observations can be used to explore these stars.
\end{enumerate}

It may be unreasonable to assume that an extraterrestrial civilization will send a signal immediately upon observing a conspicuous event. More realistically, there will be some delay between reception and transmission. Another possible reason for delay is that the civilization may wait until a time-resolved light curve of the event is complete before broadcasting a mimicked copy of the signal. For example, if the source event is a supernova (SN), they may send a transmission that will produce a light curve of the same shape and width of the supernova light curve, once the supernova luminosity has been reduced to some fraction of its maximum value. In this case, stars inside, but close to, the SETI Ellipsoid are also strong candidates for technosignature searches.

\begin{figure*}[ht]
\centering
\includegraphics[width=6in]{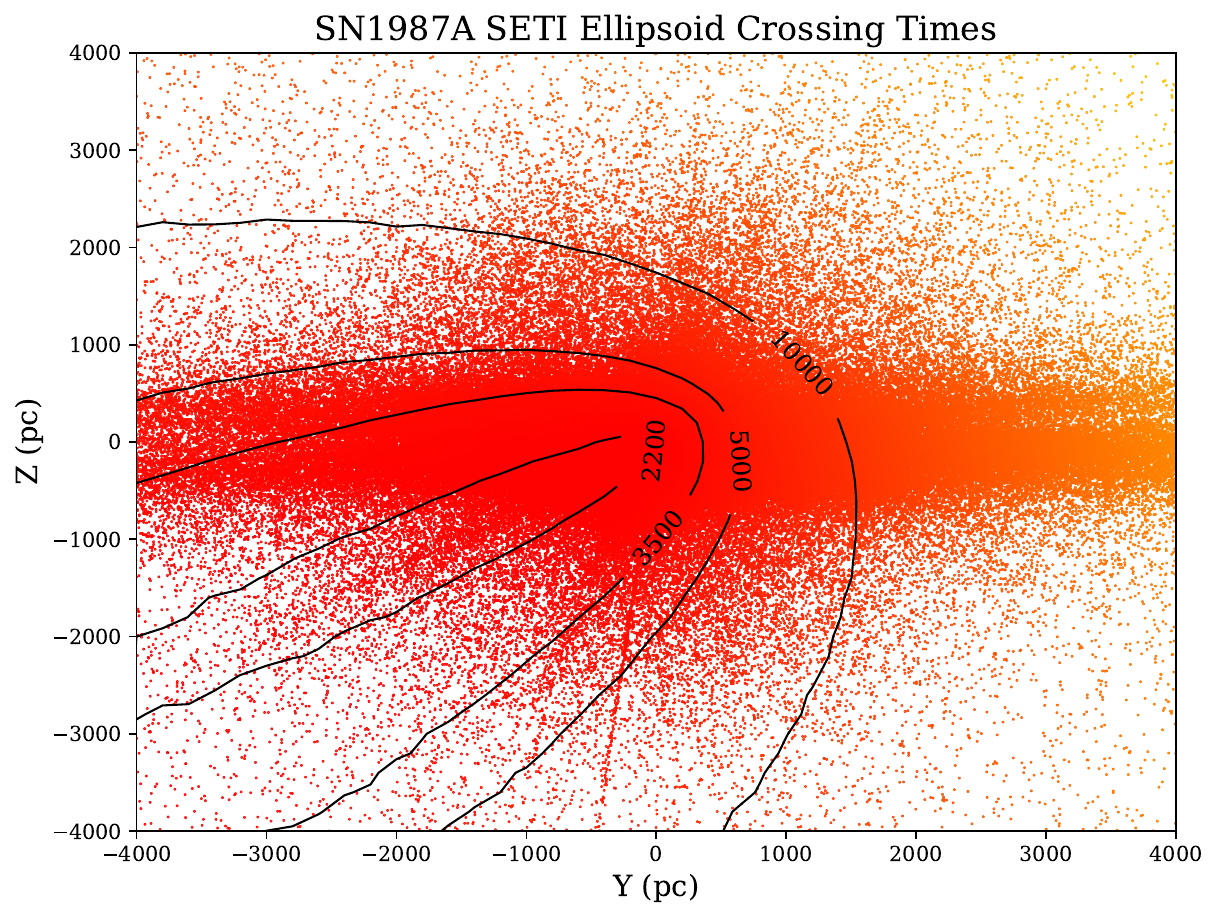}
\caption{Crossing time diagram for the SETI Ellipsoid with SN\,1987A as the source event, in galactocentric $Y$ and $Z$ coordinates, with $\lvert X \rvert < 150$ pc. The vast majority of stars will not be on the SETI Ellipsoid for thousands of years. The contour lines show the growth of the Ellipsoid.}
\label{fig:ellipsoid_crossing_times}
\end{figure*}

\begin{figure*}[ht]
\centering
\includegraphics[width=6in]{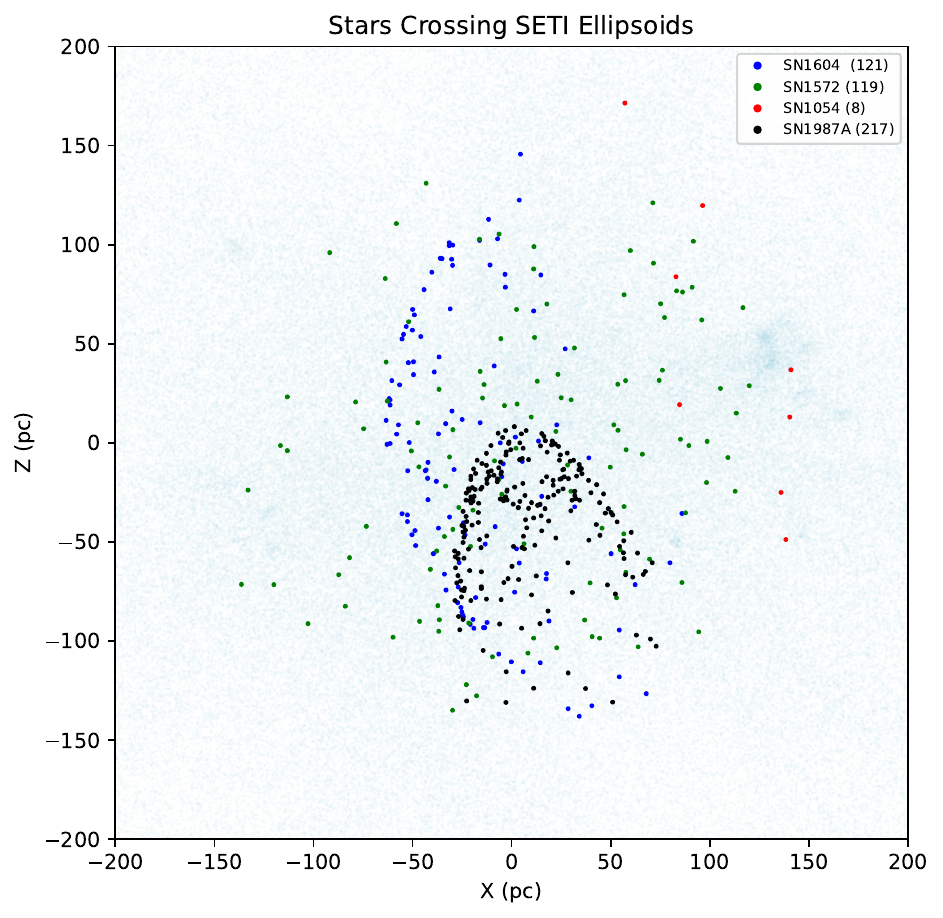}
\caption{SETI Ellipsoid target selection for SNe 1987A, 1604, 1572, and 1054, in galactocentric $X$ and $Z$ coordinates. The red dots are the 465 stars with SETI Ellipsoid crossing times, including error bounds, within the dates of available Gaia epoch photometry. The light blue dots are background stars.}
\label{fig:ellipsoid_crossings}
\end{figure*}

\begin{table*}[t]

\centering
\setlength{\tabcolsep}{1.5em}
\caption{}
Distances and Uncertainties to the Four SNe Used for Our Signaling Methods.
\begin{tabular}{c  c  c  c} 
 \hline
 SN & Distance (kpc) & Distance Uncertainty (kpc)  & Source \\ [0.5ex] 
 \hline\hline
 SN\,1987A & 51.5 & 1.2 & \citet{panagia1999} \\
 SN\,1604 & 5.1 & +0.8/-0.7 & \citet{sankrit2016} \\
 SN\,1572 & 2.75 & 0.25 & \citet{tian2011} \\
 SN\,1054 & 1.93 & 0.11 & \citet{trimble1973} \\ [1ex] 
 \hline
\end{tabular}

\label{table:sne_distances}
\end{table*}

\citet{davenport2022} focus on the SETI Ellipsoid with SN\,1987A as the source event utilizing the Gaia Catalog of Nearby Stars (GCNS), which consists of 331k stars within the 100\,pc neighborhood. SN\,1987A was chosen because of its recency and proximity, but for a Galactic signaling scheme, it may be preferable to use source events within the Milky Way. We therefore add SETI Ellipsoids with respect to SNe 1604, 1572, and 1054, which have well-known dates of observation.

Figure \ref{fig:elip} shows that the closest point on the SETI Ellipsoid to Earth is at distance $\frac{\mathscr{T}}{2} = \frac{cT}{2}$, so for source events with an age greater than about $650$\,yr, the SETI Ellipsoid will be entirely outside the 100\,pc neighborhood, and even for those with age of a few hundred years, the SETI Ellipsoid will be largely incomplete in the 100\,pc neighborhood. We thus expand our search to fully utilize the data available in Gaia DR3.

Furthermore, \citet{davenport2022} primarily consider the \textit{present} SETI Ellipsoid of SN\,1987A, which tells us only which category each star falls in, as described above. To calculate which stars are currently on the SETI Ellipsoid, we determine the distances $d_1$ and $d_2$ for each star, then use the inequality
\[\lvert d_1 + d_2 - 2\mathscr{A} \rvert \leq \tau\]
where $\tau$ is a 0.1 lyr distance tolerance of being on the Ellipsoid. The current SETI Ellipsoid for SN1987A is presented in Figure \ref{fig:1987elip}, expanded to include most of the Gaia variable star catalog. \citet{davenport2022} also calculate the crossing time for each star,
\[T_x = \frac{d_1 + d_2 - \mathcal{D}}{c}\]
which tells us the time at which the star will be on the SETI Ellipsoid, and they search through the Gaia Science Alerts archive \citep{hodgkin2021} for any variability alerts from any stars with crossing times within Gaia's operational phase. The crossing time diagram for SN\,1987A is shown in Figure \ref{fig:ellipsoid_crossing_times}. We further develop this idea by selecting only stars that have crossing times, including the error bounds, between mid-2014 and mid-2017, and then performing a variability analysis that compares the light curve before and after the crossing time. Our analysis is described in more detail in Section \ref{sec:analysis}.

The crossing time is dependent on distances to both the source event and stars, as well as the time of Earth observation of the source event, so uncertainties in these directly affect uncertainties in the time of arrival of signals to Earth. For all the SNe considered, the uncertainty in the date of observation is negligible. For small angles, the timing uncertainty is dominated by the distance uncertainty to the star, because the light travel time from the source event to Earth and the candidate star is nearly identical. This is an additional reason why \citet{davenport2022} use SN\,1987A as a source event, because its distance is large relative to nearby stars. However, this is not generally the case for Galactic SNe; the timing uncertainties for these are significantly affected by the distance uncertainties to the SNe, which are listed in Table \ref{table:sne_distances}. For many stars, these large SNe distance errors will translate to timing uncertainties that make their observation impractical, so we disregard these errors and echo the sentiment of \citet{seto2021} that we, along with several other fields of astrophysics, await high-precision distances to these objects. These are projected to become available through future projects such as the Square Kilometre Array, which will have many small dish antennas with a wider field of view that will improve parallax measurements of very radio-bright sources such as the Crab pulsar \citep{kaplan2008}, which corresponds to SN\,1054.

In total, we find 465 targets with crossing times and error bounds within the date range of available Gaia epoch photometry using the SETI Ellipsoid for the four SNe. Their spatial distribution is shown in Figure \ref{fig:ellipsoid_crossings}.


\section{Target Selection via the Seto Scheme}
\label{sec:seto}

Though the data from Gaia are a vast improvement over previous astrometric measurements, at large distances, the uncertainties in stellar distances correspond to large uncertainties in timing, which make both scheduling observations and searching through past data nonviable. To remove this error, and to complement our search of Gaia DR3 with the SETI Ellipsoid, we also use the signaling framework described by \citet{seto2021}, which does not require precise stellar distances. The basic geometry of the Seto scheme is shown in Figure \ref{fig:seto_two_angle} and summarized here.

The Seto scheme relies on finding a unique datum point in space-time that is common between Earth and an extraterrestrial civilization; in particular, the point $x$ in Figure \ref{fig:seto_two_angle}, which is the closest point to the source event on the line Ex connecting Earth and a technosignature candidate system, is such a point in space. And, the time interval $t(\theta) = l\cdot \textrm{sin}(\theta)/c$ plus the time epoch of the source event, $t_0$, is a unique temporal point. We should then search for signals that are synchronized with a transmission from the datum point $(x, t(\theta) + t_0)$. The line segment Ex in Figure \ref{fig:seto_two_angle}, with length $d = l\cdot \textrm{cos}(\theta)$ represents the depth we can search to, because any extraterrestrial civilizations on the line beyond $x$ would need to be able to predict the source event in order to send a signal that will reach Earth in synchronization with a signal from the datum point.

The time on Earth to observe each angle is given by the time of arrival of a signal sent from the datum point,
\[t_E(\theta) = (t_0 + t(\theta)) + \frac{l\cdot \textrm{cos}(\theta)}{c} - (t_0 + \frac{l}{c})\]
\[= \frac{l}{c}(\textrm{sin}(\theta) + \textrm{cos}(\theta) - 1)\] 
where $t_E$ is the time on Earth since the observation of the source event (which occurs at time $t_0 + \frac{l}{c}$). To simplify, \citet{seto2021} introduces the normalized time, 
\[\tau_\text{E} \triangleq \frac{c}{l} t_E\]
in which case the normalized times of observation are
\[\tau_\text{E}(\theta) = \textrm{sin}(\theta) + \textrm{cos}(\theta) - 1\]

\begin{figure*}[p]
\centering
\includegraphics[width=6in]{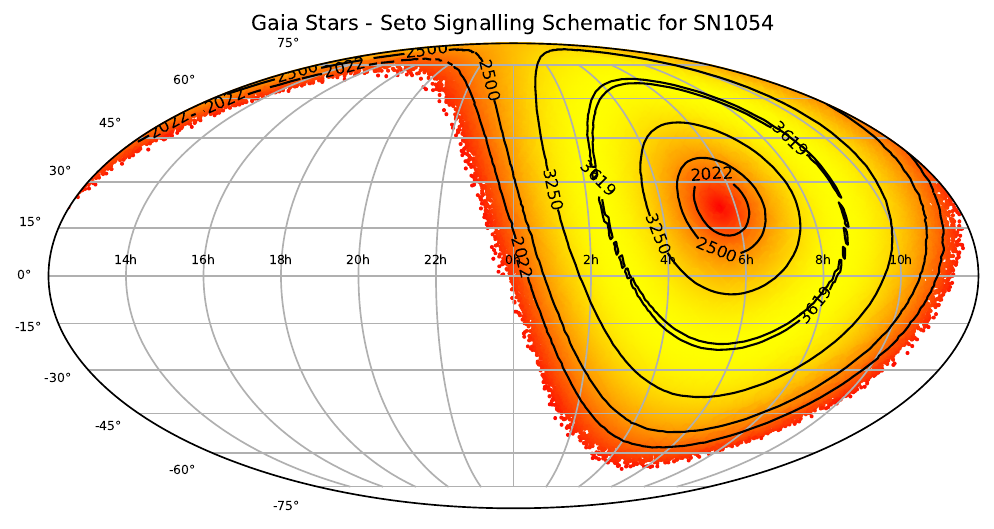}
\caption{Seto scheme crossing time diagram with SN\,1054 as the source event. This scheme covers half the sky, centered at SN\,1054, with two rings, one expanding with time and one contracting. The contour lines show the progression of these two rings. For SN\,1054, the search window closes at around the year 3620, corresponding to a normal time of $\sqrt{2}-1$.}
\label{fig:seto_crossing_times}
\end{figure*}

\begin{figure*}[p]
\centering
\includegraphics[width=6in]{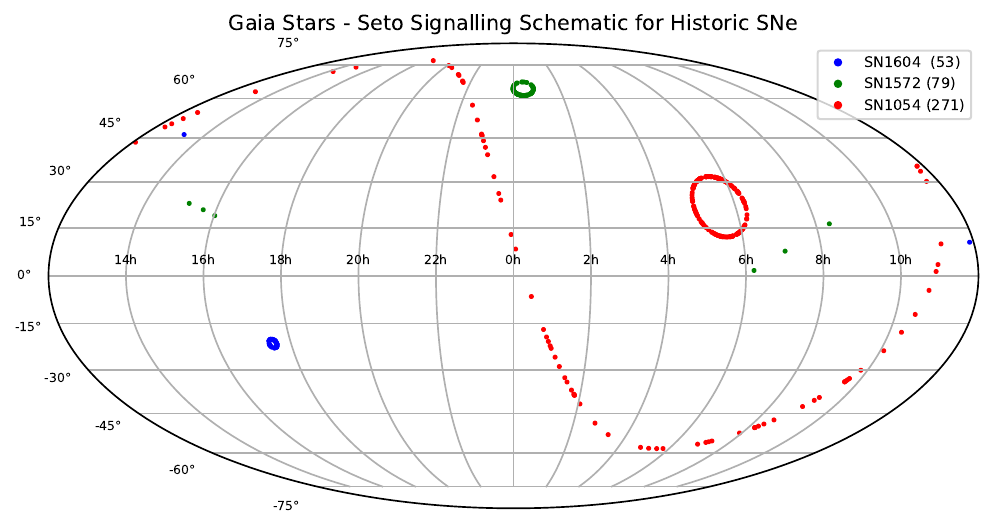}
\caption{Stars with Seto scheme crossing times within dates of available Gaia light-curve data for SNe 1604, 1572, and 1054. Each supernova has two rings of candidates, corresponding to the $\theta_+$ and $\theta_-$ rings. There are 403 candidates total.}
\label{fig:seto_gaia_crossings}
\end{figure*}

The above equation tells us the time to observe a specific angle in the sky; this can give us the crossing times of stars, directly analogous to the SETI Ellipsoid crossing times, because we can trivially calculate the angle between two sets of celestial coordinates (i.e.,\ that of the source event and that of each star). The crossing time diagram via the Seto scheme for SN\,1054 is shown in Figure \ref{fig:seto_crossing_times}. To find what angles to observe at given times, which is useful for forecasting targets of interest, we can invert the above equation. For $0 \leq \tau_\text{E} \leq \sqrt{2}-1$, there exist two solutions:
\[\theta_{\pm} = \frac{\pi}{4} \pm \textrm{cos}^{-1} \bigr(\frac{1+\tau_\text{E}}{\sqrt{2}}\bigr)\]
So, for normalized times $\tau_\text{E} < \sqrt{2}-1$, the directions to observe form two concentric rings on the celestial sphere centered at the direction of the source event. The $\theta_-$ ring begins as a point in the event direction and expands over time, while the $\theta_+$ ring begins as a great circle perpendicular to the event direction and shrinks over time; at $\tau_\text{E} = \sqrt{2}-1$, the rings merge, and beyond that, the search window of the source event has closed, and half the sky has been covered.

As noted previously, this search framework only requires that the distance to the star be less than the search depth of the respective angle, $d = l\cdot \textrm{cos}(\theta)$. At $\tau_\text{E} = 0$, the $\theta_-$ ring has a search depth of $l$ and the $\theta_+$ ring has a search depth of $0$; these values decrease and increase, respectively, to a final value of $\frac{\sqrt{2}l}{2}$.

Though not essential for a civilization that is only searching, the signaling schematic is similarly time-dependent. The directions that an extraterrestrial civilization following the Seto framework must transmit in are identical to the searching directions, but reflected across the plane perpendicular to the source event direction. In other words, the two signaling rings have the same angles as the receiving rings, but are concentric about the direction antipodal to the event direction.

A strong connection exists between the SETI Ellipsoid and the Seto scheme. In particular, the two search directions in the Seto scheme correspond to the angles at which the lines $d_1$ and $d_2$ in Figure \ref{fig:elip} are perpendicular, and the search depth is the length $d_2$. This can be seen through the equation
\[t_E(\theta) = \frac{l}{c}(\textrm{sin}(\theta) + \textrm{cos}(\theta) - 1) = \frac{d_1 + d_2 - l}{c}\]
which is identical to the crossing time equation for the SETI Ellipsoid. As seen in Figure \ref{fig:seto_two_angle}, two such angles exist, corresponding to $\theta_+$ and $\theta_-$, which form two rings in the celestial sphere when rotated about the Earth-source event axis. The normalized time $\tau_\text{E} = \sqrt{2}-1$ corresponds to the point at which no such perpendicular lines exist in the SETI Ellipsoid.

Although stellar distance uncertainties do not factor into the timing uncertainties using the Seto scheme, distance errors to the SNe will have an effect. The timing uncertainty is 
\[dt = \partial t_\theta + \partial t_l\]
This is strongly dominated by the $\partial t_l$ term, because the angular positions of the stars and SNe have extremely precise values, so $d\theta$ is small. So,
\[dt = \frac{dt_E}{dl} dl = \frac{dl(\textrm{sin}(\theta) + \textrm{cos}(\theta) - 1)}{c}\]
which has a maximum value of $dl\frac{\sqrt{2}-1}{c}$ in the domain of $\theta$. Hence, precise measurements of SNe distances are necessary for constraining crossing times to within a reasonable interval.

However, as noted in Section \ref{sec:elip}, the distances to supernova remnants have large uncertainties corresponding to unfeasibly large timing uncertainties for most angles. Thus, we again disregard these errors.

\begin{figure*}[p]
\centering
\includegraphics[width=6in]{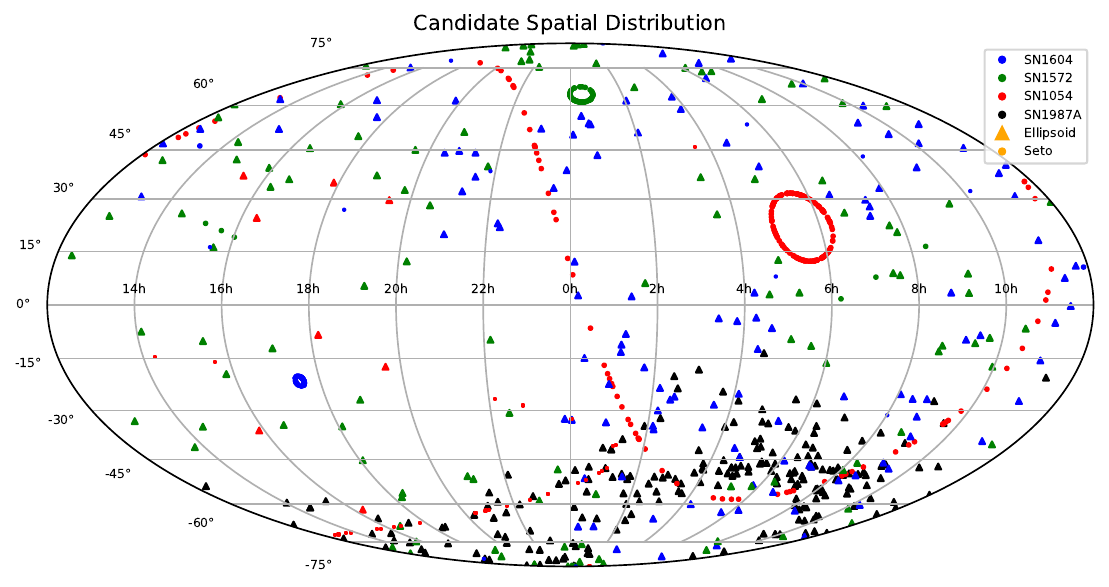}\\
\includegraphics[width=6in]{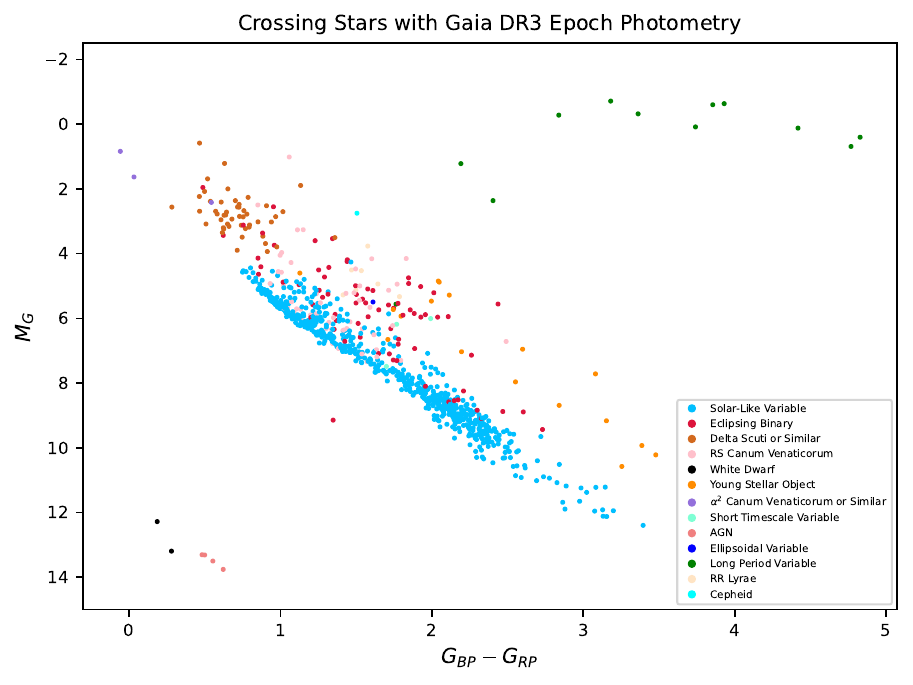}
\caption{The top panel shows the spatial distribution of all 868 candidates selected via the SETI Ellipsoid and Seto methods for all four SNe, with crossing times within the date range of available Gaia light curves, combining the targets from Figures \ref{fig:seto_gaia_crossings} and \ref{fig:ellipsoid_crossings}. The bottom panel displays the color-magnitude diagram of these targets, each marked with its Gaia variable type classification.}
\label{fig:gaia_CMD_Crossings}
\end{figure*}

For our Seto scheme target selection, we again choose stars with crossing times within the dates of Gaia epoch photometry, excluding error bounds as noted above, and with distances less than the search depths. We use SNe 1604, 1572, and 1054 as source events. SN\,1987A is excluded because its normalized time is very small, so the $\theta_+$ ring has a very low depth and the $\theta_-$ ring is very thin; when calculated, there are zero stars for the Seto scheme with respect to SN\,1987A within the desired date range. In total, we find 403 targets with crossing times within the date range of available Gaia epoch photometry using the Seto method for the three SNe. Their spatial distribution is shown in Figure \ref{fig:seto_gaia_crossings}.

\section{Candidate Exploration}
\label{sec:cand}

We find a total of 868 candidates with SETI Ellipsoid or Seto crossing times falling in the time range of Gaia DR3 epoch photometry. Their astrometric properties, crossing times, and corresponding source supernovae are listed in Table \ref{table:candidatelist}, which is available in full in the machine-readable format. The spatial distribution and color-magnitude diagram, with Gaia variability classification, of the target stars is shown in Figure \ref{fig:gaia_CMD_Crossings}. 

The majority of candidates are classified as solar-like variables, which have rotational modulation, flares, and spots. These are more difficult to analyze, because they have a less well-defined model that involves the many parameters of the star's magnetic active regions. However, a possible advantage is their similarity to the Sun, which may indicate a greater chance at hosting life compared with other variable types, many of which clearly lie above the main sequence, but we do not eliminate the possibility of extraterrestrial civilizations with stars of different evolutionary states or with more extreme variability.

Although variable stars are astrophysically interesting, and we expect a form of artificial variability when searching for extraterrestrial intelligence, it is unclear whether or not any type of technosignature signal in a stellar light curve would be classified as a variable star by Gaia's machine-learning classification algorithm \citep{gaiadr3variability}. Future Gaia data releases will include epoch photometry for all sources in the catalog, allowing for a more complete analysis. However, we note that the classifications from Gaia are only the most likely variable type for each star, outputted from the machine-learning algorithm, and so the stars may be classified incorrectly, may not fall within any of Gaia's predefined variability types, or may not even be

\startlongtable
\movetabledown=2.75in

\begin{rotatetable*}

\begin{deluxetable*}{ c  c  c  c  c  c  c  c  c  c  c  c  c  c  c  c  c }
\tabletypesize{\scriptsize}
\tablecaption{Table of Candidate Targets\label{table:candidatelist}}

\tablehead{\colhead{ID} & \colhead{RA} & \colhead{Dec} & \colhead{Dist\tablenotemark{a}} & \colhead{Dist84\tablenotemark{a}} & \colhead{Dist16\tablenotemark{a}} & \colhead{X\tablenotemark{b}} & \colhead{Y\tablenotemark{b}} & \colhead{Z\tablenotemark{b}} & \colhead{$G$\tablenotemark{c}} & \colhead{$G_{BP}$\tablenotemark{c}} & \colhead{$G_{RP}$\tablenotemark{c}} & \colhead{Class\tablenotemark{d}} & \colhead{XTime\tablenotemark{e}} & \colhead{Seto\tablenotemark{f}} & \colhead{SN} & \colhead{XTimeErr\tablenotemark{e}} \\  & \colhead{$^\circ$} & \colhead{$^\circ$} & \colhead{pc} & \colhead{pc} & \colhead{pc} & \colhead{pc} & \colhead{pc} & \colhead{pc} &  &  &  &  & \colhead{BJD} &  & & \colhead{yr}}

\startdata
 5277882523178810112 & 126.44 & -61.89 & 166.7108 & 166.95 & 166.42 & 18.46  & -161.00 & -39.15 & 12.31 & 12.81 & 11.65 & SOLAR\_LIKE  & 2457250.23 & False & 1987A & 0.85 \\
 5278211586393119104 & 125.85 & -60.68 & 160.97 & 161.21 & 160.70 & 14.58 &   -156.07 & -36.63 & 13.35 & 14.04 & 12.54 & SOLAR\_LIKE & 2457562.62 & False & 1987A  & 0.83 \\  
 5085376179094311680 & 55.70 & -24.46 & 26.00 & 26.01 & 25.98 & -12.62 &  -10.15 & -20.32 & 14.03 & 15.83 & 12.76 & SOLAR\_LIKE & 2457030.67 & False & 1987A & 0.04 \\
 5882042344287383936 & 232.72 & -58.86 & 25.15 & 25.16 & 25.14 & 19.91 &  -15.37 & -0.98 & 10.20 & 11.18 & 9.21 & SOLAR\_LIKE & 2457524.21 & False & 1987A & 0.04 \\ 
 5793290899490805888 & 220.68 & -74.31 & 54.03 & 54.08 & 54.00 & 34.09 & -40.10 & -12.31 & 8.30 & 8.64 & 7.78 & ECL & 2457762.37 & False & 1987A & 0.13 \\ 
 5573439907376720896 & 95.79 & -40.22 &  66.51 & 66.56 & 66.46 & -23.11 &  -57.09 & -25.09 & 10.62 & 11.19 & 9.90 & SOLAR\_LIKE & 2457294.68 & False & 1987A & 0.16 \\
 5575165556516524416 & 96.17 & -36.67 & 51.70 & 51.74 & 51.67 & -20.84 &  -43.61 & -18.33 & 12.99 & 14.28 & 11.87 & SOLAR\_LIKE & 2456961.20 & False & 1987A & 0.11 \\ 
 5764995891157361792 & 203.08 & -88.44 & 128.03 & 128.22 & 127.87 & 63.18 & -96.57 & -55.47 & 11.49 & 11.92 & 10.88 & SOLAR\_LIKE & 2457519.76 & False & 1987A & 0.57 \\
 5765214766988828672 & 268.90 & -87.94 & 109.54 & 109.66 & 109.41 & 56.30 & -80.06 & -49.22 & 13.68 & 14.65 & 12.70 & SOLAR\_LIKE & 2457048.89 & False & 1987A & 0.40 \\
 5258643234376355840 & 153.30 & -57.82 & 63.07 & 63.12 & 63.03 & 14.31 & -61.42 & -1.33 & 13.19 & 14.31 & 12.14 & SOLAR\_LIKE & 2457547.73 & False & 1987A & 0.16 \\
\enddata

\tablenotetext{a}{Distances and distance uncertainties (upper and lower bounds at the 84th and 16th percentiles, respectively), are taken from \citet{bailer-jones2021}}
\tablenotetext{b}{Galactocentric Cartesian $X$, $Y$, and $Z$ coordinates.}
\tablenotetext{c}{Gaia magnitudes in $G$, $G_\text{BP}$, and $G_\text{RP}$ bands.}
\tablenotetext{d}{Gaia variability type classification.}
\tablenotetext{e}{Star crossing time via either the SETI Ellipsoid or the Seto method.}
\tablenotetext{f}{True if the star is a candidate via the Seto method, and false if the star is a candidate via the SETI Ellipsoid.}
\tablecomments{Table 2 is published in its entirety in the machine-readable format. A portion is shown here, rounded to two decimals, for guidance regarding its form and content.}

\end{deluxetable*}
\end{rotatetable*}

\noindent variable stars at all, and so a technosignature search in the light curves of these objects is not futile. Our light-curve analysis described in Section \ref{sec:analysis} is sensitive to both changes in variability parameters and flux, and so can be applied to both variable and nonvariable stars, regardless of their Gaia classification. Furthermore, it can be modified given prior information about the sample of stars to be analyzed; for example, we perform a more detailed analysis for the sample of eclipsing binaries, which comprise a significant fraction of the candidates, as seen in Figure \ref{fig:gaia_CMD_Crossings}.

The distance errors for the 465 SETI Ellipsoid candidates must be less than 1.5\,ly, because otherwise the timing uncertainties would necessarily extend outside the Gaia data window, which spans three years. The actual maximum error is 0.46\,ly, and 68\% of candidates have errors less than 0.19\,ly, indicating that for the majority of these targets, the crossing times are accurate to approximately two months.

Predictably, the distance errors for the 403 Seto method candidates are much greater, as they have no constraints apart from the upper bound distance being less than the search depth. For these targets, the maximum error is 2000\,ly, and 66\% have errors less than 55\,ly.

\section{Variability Analysis}
\label{sec:analysis}

\begin{figure*}[t]
\centering
\includegraphics[width=6in]{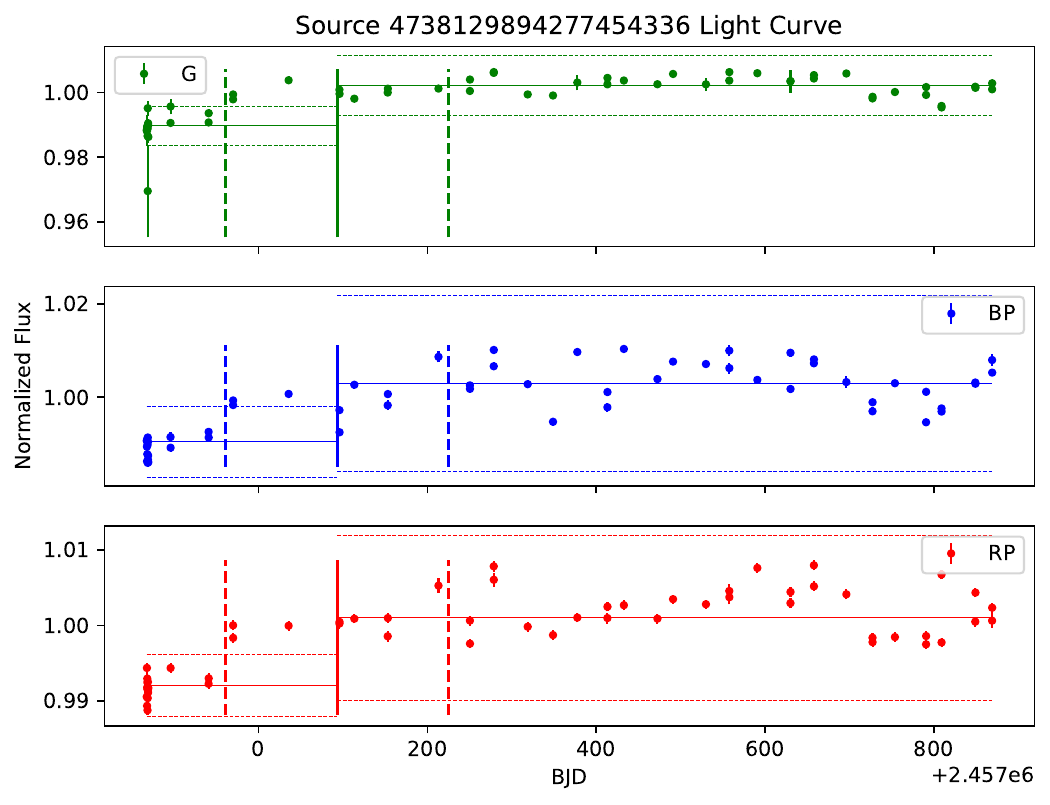}
\caption{Sample light curve for one of the selected targets, with solar-like variable type classification, in the three Gaia bands: $G$, $BP$, and $RP$. The solid vertical line marks the crossing time, via either the SETI Ellipsoid or the Seto scheme. The solid horizontal lines denote the median flux, calculated separately before and after the crossing time, and the dashed horizontal lines are the $\pm 3 \times \textrm{median absolute deviation (MAD)}$ levels. This target is a candidate from the SETI Ellipsoid method, so we include the crossing time error due to the stellar distance uncertainty, which is denoted by the vertical dashed lines. The data are noticeably sparse and incomplete, consisting of only three years of Gaia observations, but have long-term stability.}
\label{fig:sample_LC}
\end{figure*}

To fully utilize the available epoch photometry data from Gaia (see Figure \ref{fig:sample_LC}), which is limited by sparsity and incompleteness, we use a novel technosignature search approach that splits the light curves at the crossing time and looks for variations between the two halves. We focus on periodic variables, for which the sparsity issue can be alleviated by period folding.

In particular, eclipsing binaries are the most ideal variable type for our analysis, because they have easily parametrized light curves as double Gaussians. This allows for a numerical measure of the dissimilarity of the left and right halves of the light curve using an error-weighted difference for each parameter. Ranking these measures can then give the most anomalous light curve, which would correspond to the light curve with the greatest change at the crossing time and thus the highest potential to be a technosignature system.

\begin{figure*}[t]
\centering
\includegraphics[width=6.5in]{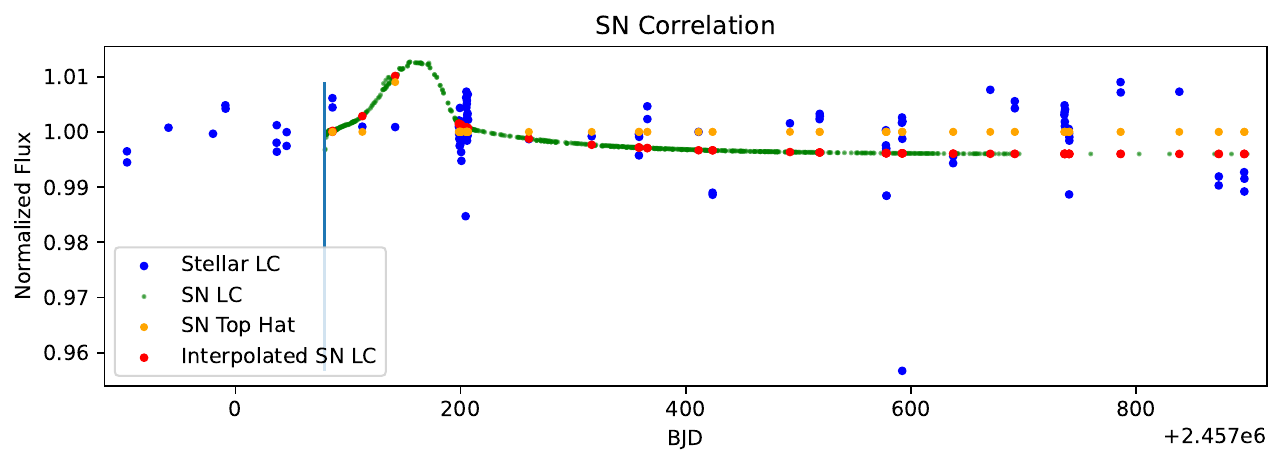}
\caption{Sample candidate light curve with SN\,1987A light curve overlain in green. The red dots show the SN light-curve interpolated to match the Gaia light curve data points; the orange dots are the same for a top-hat function with width equal to the FWHM of the SN light curve, a baseline equal to a normalized flux of 1, and a peak equal to the maximum normalized flux of the stellar light curve. The vertical blue line indicates the crossing time of the star, which is where the supernova light curve and top-hat function begin. The timescale of the SN, on the order of hundreds of days, is close to that of the Gaia epoch photometry sampling and is also longer than the periods of most stars that are classified as variable by Gaia. This approach would be better suited to a dataset that has denser time sampling and is not limited to only variable stars.}
\label{fig:sn_xcorr}
\end{figure*}

Here we point out that rather than a traditional SETI approach, which generally looks for direct electromagnetic emission from technology with spectrotemporal properties consistent with known or hypothesized behaviors of technology, our method is sensitive to civilizations that can modulate their host star's period, amplitude, or phase. While this may be possible for a sufficiently advanced civilization, we ideally would like to search for several types of potential transmissions, such as a signal mimicking the light curve of the source event, across all stars, not just those classified as periodic or variable. We attempted a search for such signals by cross-correlating a normalized supernova light curve, as well as a simpler top-hat function derived from the same parameters of the supernova light curve, with each candidate's light curve; however, the Gaia epoch photometry is too sparse for this analysis, as the timescale of a supernova is on the same order as the spacing between photometry measurements, making any correlation spurious. This can be seen in Figure \ref{fig:sn_xcorr}.

Of the 868 candidate targets, 73 are classified as eclipsing binaries. For each of these eclipsing binaries, we first split the normalized $G$-band light curve at the crossing time. We refer to the light curves before and after the crossing time as the left and right light curves, respectively. Then, with a Lomb-Scargle Periodogram, we extract the peak frequency of the full, left, and right light curves. Each of these frequencies also carries a false alarm probability, which is the probability that, assuming the light curve has no periodic signal, we will observe a periodogram power at least as high; we take these values to be the error, although they do not directly represent a statistical uncertainty.

All three light curves are subsequently folded with the same frequency, namely the peak frequency with the lowest false alarm probability, and in reference to the same epoch. A fit to a double Gaussian with constant baseline is then performed for each folded light curve, and seven relevant parameters are extracted: the depth, phase, and width of the two Gaussians, and the baseline flux. For these parameters, plus the median flux and the peak frequency, we calculate the error-weighted difference,
\[\frac{\lvert X_\textrm{right} - X_\textrm{left} \rvert}{\sqrt{\sigma_\textrm{right}^2 + \sigma_\textrm{left}^2}}\]
where $\sigma$ is the square root of the variance for the seven double Gaussian parameters, the MAD for the median flux, and the false alarm probability for the peak frequency.

For many of these targets, the crossing time is close to either the start or the end of available observations, which causes either the left or right light curve to contain very few data points. In these cases, we are unable to perform a double Gaussian fit at all. Thus, from the 73 eclipsing binaries, we ultimately perform a full analysis on 45 of them. The analysis was performed solely on $G$-band data, because we find that trying to fit all three bands simultaneously further reduces the number of successful fits; however, if a promising signal is found from the $G$-band data, it would be valuable to analyze the $RP$ and $BP$ bands to confirm that the signal is present in them as well.

For each parameter, we rank the targets in order of increasing error-weighted difference, so the $0^\textrm{th}$ target has the most similar left and right light curves with respect to a given parameter, and the $44^\textrm{th}$ target has the most dissimilar light curves. We then take the sum of the rankings for each target and divide by the maximum, leaving us with a single interest index ranging from 0 to 1, where 1 represents the most interesting target (i.e.,\ the target with the greatest difference between its left and right light curves).

\begin{figure*}[p]
\centering

    \includegraphics[width=3.5in]{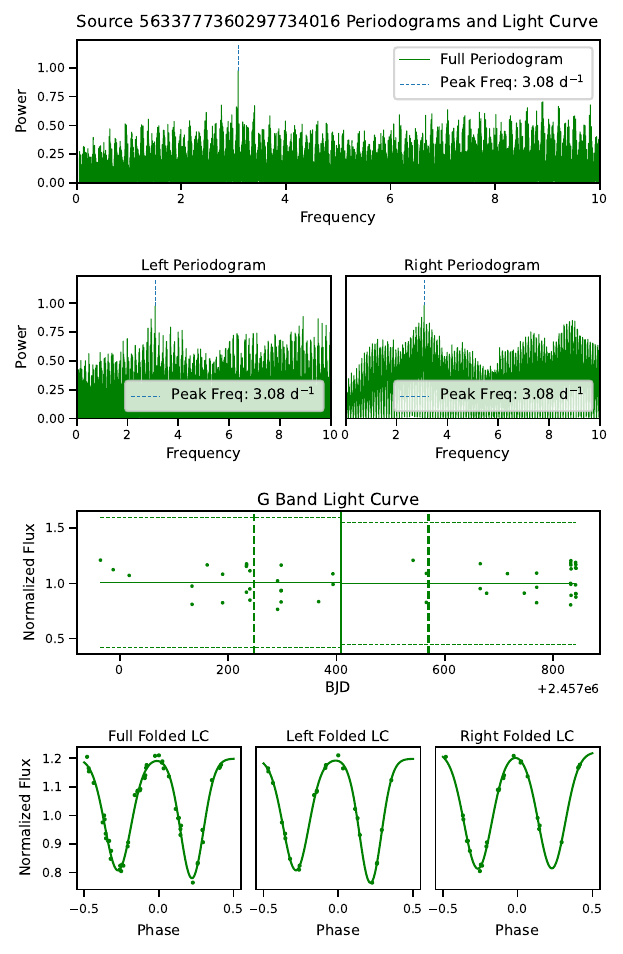}
    \includegraphics[width=3.5in]{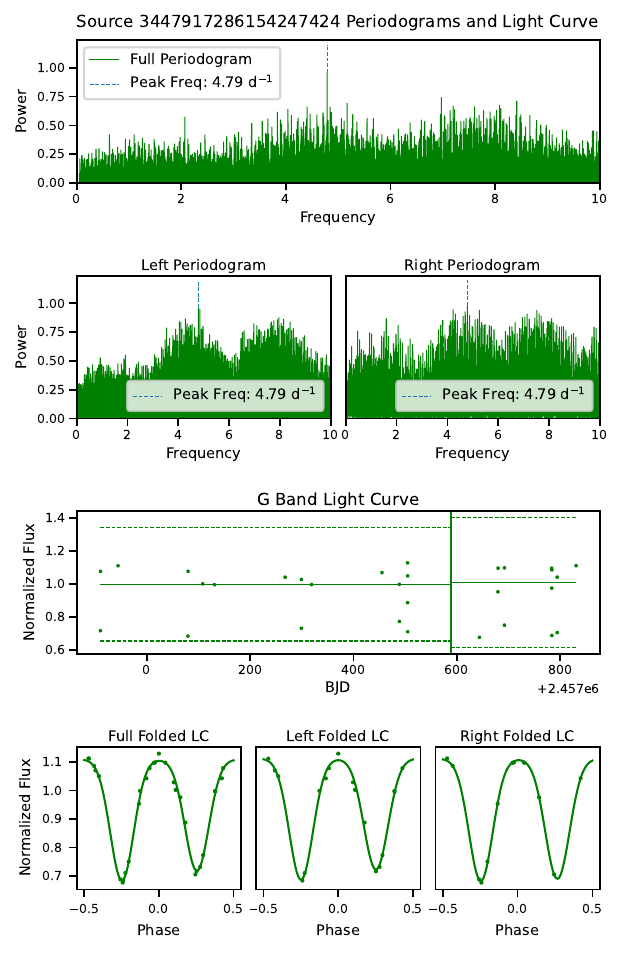}
    
    \caption{(a) Full, left, and right light curves and their Lomb-Scargle Periodograms for two eclipsing binaries. The left panel has the highest-interest index and the right panel is near the middle of the interest rankings. The analysis seems to successfully fold the data both before and after the crossing time, and there is a small but noticeable change in the double Gaussian eclipsing binary fit between the left and right light curves, though this is most likely due to sparse sampling. It is difficult to distinguish between eclipsing binaries with high interest, as we do not normally expect the variability parameters for an eclipsing binary to change, so the left and right light curves should be similar. (b) Lomb-Scargle Periodograms and light curves for two eclipsing binaries with some of the lowest-interest indices. The periodograms are unable to find strong peak frequencies and the double Gaussian fit fails in these two panels, indicating that the analysis and ranking system successfully assigns low interest to targets that do not have clean data.}
    \label{fig:eclipsing_binaries}
    
\end{figure*}

\begin{figure*}[t]
\ContinuedFloat\centering

    \includegraphics[width=3.5in]{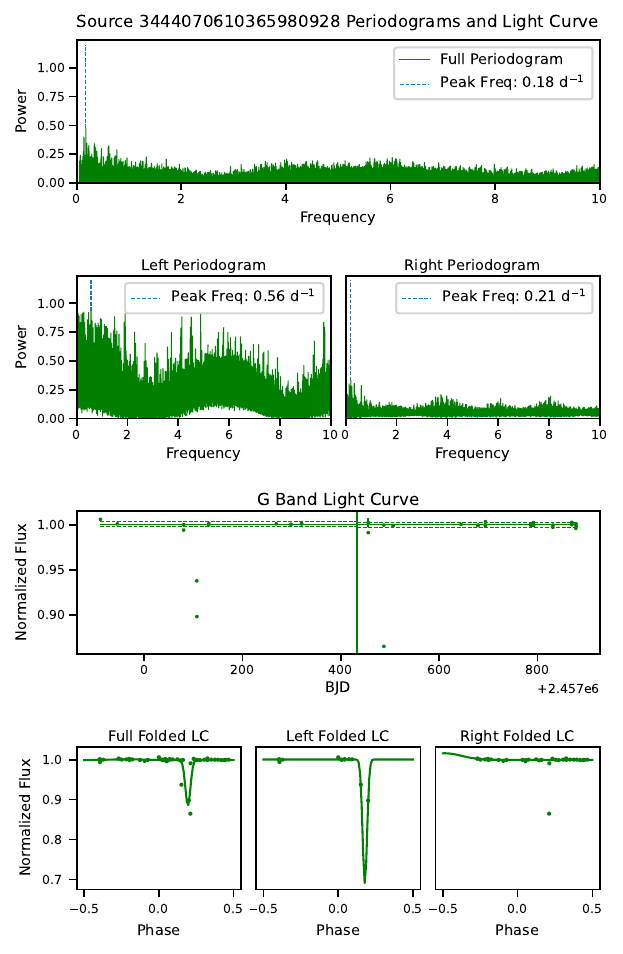}
    \includegraphics[width=3.5in]{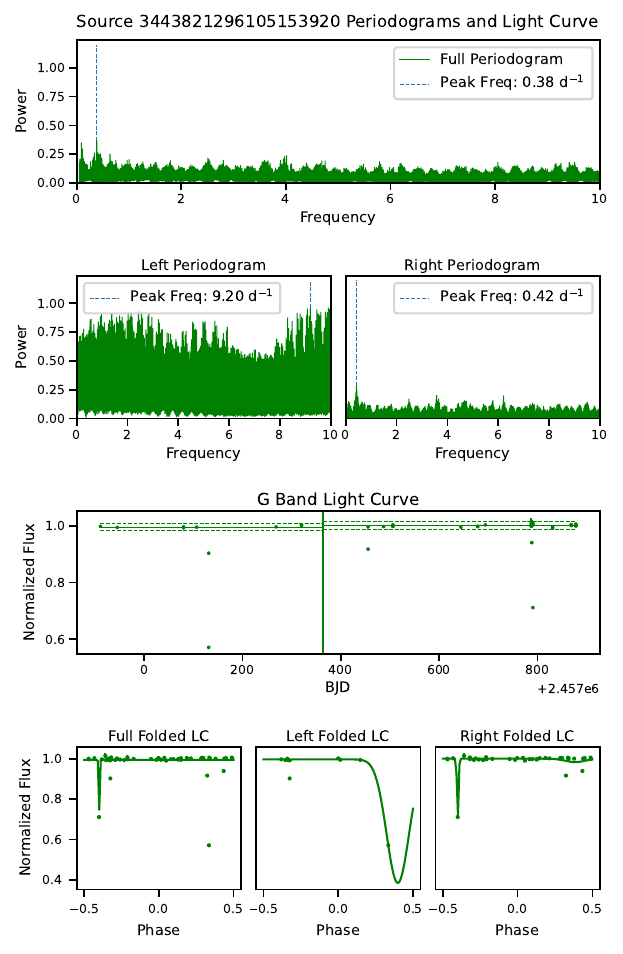}
    \caption{(Continued.)}
\end{figure*}

The periodograms and light curves for four eclipsing binaries, including the target with the highest-interest index, are shown in Figure \ref{fig:eclipsing_binaries}. There is not a large difference between targets of high interest, as seen in the top two panels, but this is expected given the null hypothesis of no signals. In the most interesting candidate (top left panel), the greatest difference between the left and right light curves is due to a change in the relative depths of the two Gaussians. However, upon inspection, we see that this difference, and likely most other changes between the left and right light curves for these systems, is caused by the sparsity of the light curve, even when folded.

However, our variability analysis and ranking algorithm are successful in separating eclipsing binaries with clearly distinctive light curves from those without. The lowest-ranked light curves, in the bottom panels of Figure \ref{fig:eclipsing_binaries}, clearly have much worse Lomb-Scargle Periodogram frequency extractions and double Gaussian fits, largely because there are very few points in the light-curve dips.

Although a full periodicity analysis cannot be implemented on the entire sample of variable stars, because many are not periodic, we perform a similar but simpler analysis on the remaining candidates. Rather than comparing the double Gaussian fit parameters before and after the crossing time, we compare the median flux, the standard deviation, and the best-fit slope for these targets. We follow the same process of calculating the error-weighted differences for the median flux and best-fit slope, but for the standard deviation, we instead normalize by the mean flux and calculate a simple difference, because there is no well-defined measure of the error of a standard deviation value. An advantage of this simpler procedure is that we can incorporate all three Gaia filter bands, because a linear fit will fail only if there are fewer than two points on either side of the crossing time, so we ultimately compute nine parameters, the same number as for the eclipsing binary analysis, for a total of 734 out of the 795 variable stars not classified as eclipsing binaries.

\begin{figure*}[p]
\centering
\includegraphics[width=5in]{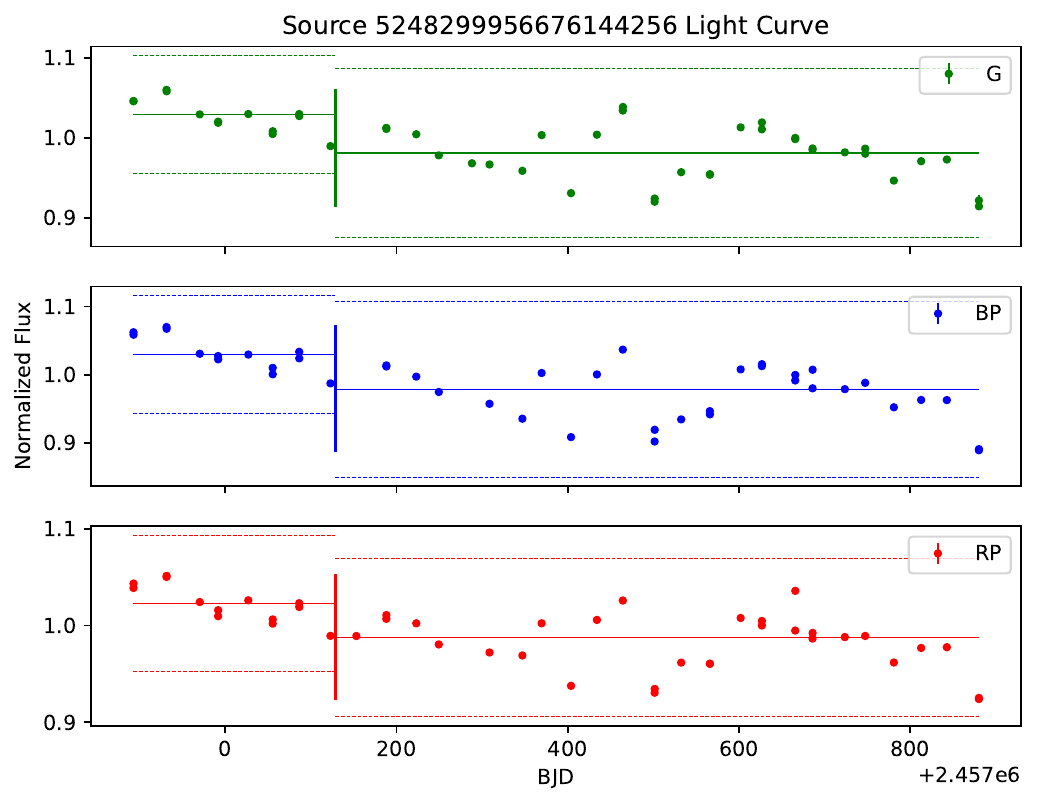}\\
\includegraphics[width=5in]{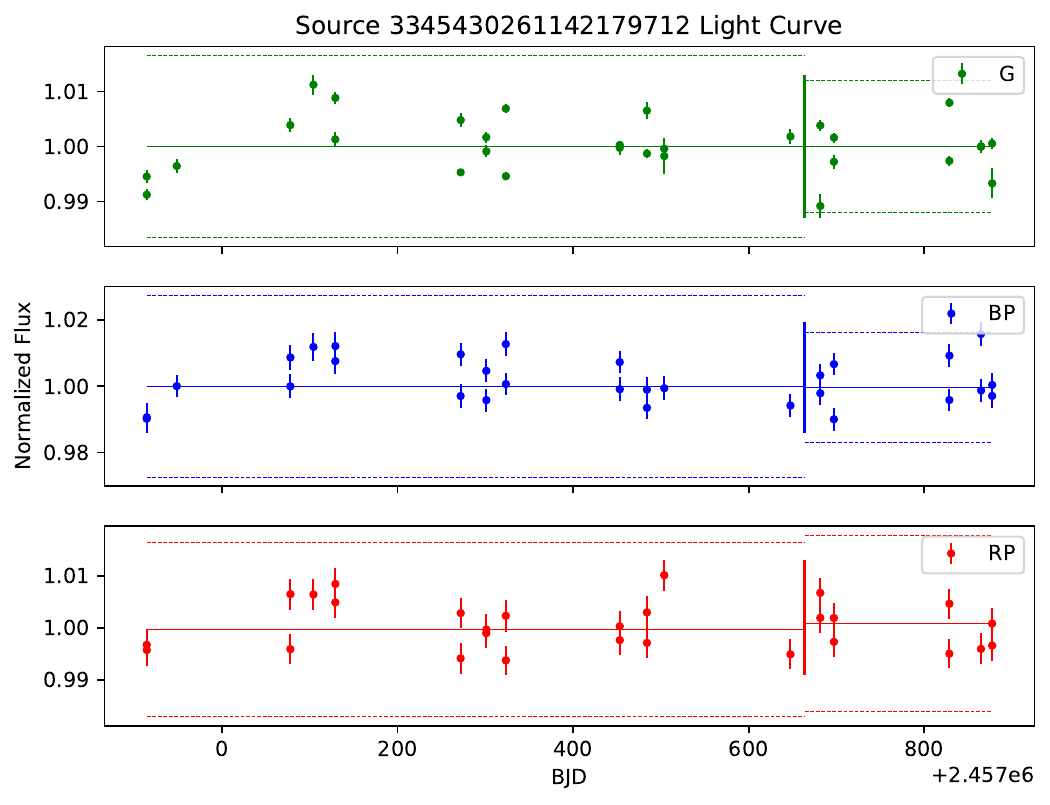}
\caption{Light curves in all three Gaia bands for the most (top panel) and least (bottom panel) interesting targets not classified as variable stars, formatted in the same manner as in Figure \ref{fig:sample_LC}. The most interesting candidate displays a nonsignificant but noticeable shift in the median flux before and after the crossing time, as well as possible increase in the scatter. The best-fit slope also appears to be strongly negative before the crossing time but near-horizontal afterward. On the other hand, the least interesting candidate shows almost zero change in the median flux, and seemingly small change in the scatter and best-fit slope.}
\label{fig:non_eclipsing_binaries}
\end{figure*}

We also repeat the same target ranking process, and show the most and least interesting candidates in Figure \ref{fig:non_eclipsing_binaries}. It again appears that our analysis successfully finds uninteresting candidates with seemingly well-understood or periodic light curves, giving them low rankings and separating them from more interesting candidates, which appear to have more significant changes in light curves before and after the crossing time. However, it remains difficult to separate any possible technosignature signal from random variation and noise given the limits of the Gaia epoch photometry.

\section{Conclusions}
\label{sec:conclusion}

We have presented both a spatiotemporal technosignature candidate search framework, which combines the SETI Ellipsoid and Seto methods, and a novel SETI approach that is sensitive to changes in a periodic star's variability parameters. With precise astrometry and distance from Gaia and \citet{bailer-jones2021}, we explore this framework with several nearby SNe as source events for signal synchronization to select candidates with crossing times within the time range of available Gaia epoch photometry, and perform our variability analysis on the well-parametrized subset of eclipsing binary candidate systems.

Given the lack of any statistically significant changes in the light curves of the selected eclipsing binaries, we can place constraints on the prevalence of extraterrestrial civilizations using either of these spatiotemporal signaling schemes and behaving in a manner detectable by our time domain analysis, as has been done by \citet{price2020}, \citet{franz2022}, and others for previous SETI observations. This limit can be calculated with a one-sided 95\% Poisson confidence interval, assuming a conservative 50\% probability of detecting such a signal if present \citep{gehrels1986}. We performed our analysis on a total of 779 variable star systems, and so we place an upper limit of 0.47\% on the percentage of such systems hosting civilizations that are behaving in the hypothesized manner.

Although we use only the error associated with the distance to the stars to calculate crossing time uncertainties, the errors in the SNe distances are also significant, and will have a nonnegligible effect on the timing uncertainty for most targets via both the SETI Ellipsoid and the Seto method. We, like \citet{seto2021}, again emphasize that while precise distances to SNe will help us more accurately constrain our technosignature searches, the advent of these measurements is beneficial not just to SETI but to many areas of astrophysics. With tighter SNe distance errors, we will also be able to expand our framework to other historic Galactic SNe.

Our geometric framework constrains the SETI parameter space by telling us where and when to search, for both planned observations and archival data, but the nature of the synchronized transmission is still unknown. To best utilize the light curves from Gaia, we look for modulations in frequency, amplitude, or phase of the candidate systems, which is reasonably feasible for an advanced civilization. Another possible intentional signals in optical bands is a peak in the light curve that mimics the shape and width of the source event. Although this is not yet feasible given the limitations of Gaia data and photometry, the situation may change in the near future. This method may also be better suited for a telescope like TESS, which has more densely sampled data at the cost of some long-term stability and discontinuous observations for most stars. Furthermore, there is the possibility of using other datasets and observatories to search for synchronized bright radio flashes or other types of transient radio signals. In particular, the SETI Ellipsoid and Seto schemes could be used to select targets for radio SETI observations.

This work is also limited by the incompleteness of Gaia DR3 light curve data, which consist of a curated set of variable objects that have passed through a classification algorithm and may not include other potentially interesting candidates, such as stars that may have a short, small increase in flux. Moreover, only three years of photometric data have been released for the stars with light curves. Gaia has made this work and other geometric approaches to SETI possible, and we expect that the next big revolution will arrive with Gaia Data Release 4, which will include 66 months of epoch photometry for all 2 billion sources.

\begin{acknowledgements}
The authors thank the anonymous referee for their helpful comments, which substantially improved this work.
The authors wish to thank Sofia Sheikh and Barbara Cabrales for helpful conversations about the SETI Ellipsoid.

The Breakthrough Prize Foundation funds the Breakthrough Initiatives which manages Breakthrough Listen. A.N. was funded as a participant in the Berkeley SETI Research Center Research Experience for Undergraduates Site, supported by the National Science Foundation under grant No.~1950897.

J.R.A.D. acknowledges support from the DiRAC Institute in the Department of Astronomy at the University of Washington. The DiRAC Institute is supported through generous gifts from the Charles and Lisa Simonyi Fund for Arts and Sciences, and the Washington Research Foundation.

\end{acknowledgements}

\software{
Python, IPython \citep{ipython}, 
NumPy \citep{numpy}, 
Matplotlib \citep{matplotlib}, 
SciPy \citep{scipy}, 
Pandas \citep{pandas}, 
Astropy \citep{astropy}}

\bibliography{references.bib}
\bibliographystyle{aasjournal}

\end{document}